\documentstyle[11pt]{article}

\def\appendix{\par
 \setcounter{section}{0}
 \setcounter{subsection}{0}
 \def\thesection{Appendix \Alph{section}}
 \def\theequation{\Alph{section}.\arabic{equation}}
 \setcounter{equation}{0}}

\begin{document}

\begin{flushright}
DO-TH 95/20, THES-TP 95/14 \\
(hep-ph/9512333)
\end{flushright}
\centerline{\Large \bf Goldstone modes and the Higgs condensation} 
\centerline{\Large \bf beyond one loop} \vspace{1.2cm}

\begin{center}
{{\bf G.~Cveti\v c}\\Inst.~f\"ur Physik, Universit\"at Dortmund, 
44221 Dortmund, Germany\\
[0.8cm]{\bf N.D.~Vlachos}\\Dept.~of Theor.~Physics, Aristotle 
University of Thessaloniki, 540 06 Thessaloniki, Greece }
\end{center}

\vspace{1.2cm} \centerline{\bf Abstract} We study the Higgs 
condensation $H=\langle \bar tt\rangle $ mechanism in the 
Top-mode Standard Model at the next-to-leading order in 
$1/N_{\mbox{\footnotesize c}}$. 
The calculation includes the effects of the Goldstone fields,
but not the effects of the transverse components of the
electroweak gauge bosons. The 
resulting effective theory is parametrized by means of a finite 
energy cut-off $\Lambda $ at which the condensation is supposed to 
take place. Demanding that the next-to-leading order contributions 
not dominate over the leading order ones, we get a rather low bound 
for the cut-off: $\Lambda = {\cal {O}}(1\mbox{TeV})$. QCD effects 
can change the results somewhat, but the basic conclusions remain 
unchanged. The inclusion of the Goldstone degrees of freedom tends 
to decrease the bound on $\Lambda $ .\\
PACS number(s): 12.60.Rc, 14.80.Bn, 11.15.Pg

\section{Introduction}

The idea that the Higgs mesons could be bound states of heavy quark 
pairs has been developed and worked on in a series of papers by 
various authors (~\cite{nambu}-~\cite{justin}, and references 
therein), motivated by an earlier work of Nambu and 
Jona-Lasinio (NJL)~\cite{njl}. The bound states (condensates) are 
treated in these works either in the 
leading-$N_{\mbox{\footnotesize c}}$ approximation, or in a form 
that takes into account part of the effects beyond the 
leading-$N_{\mbox{\footnotesize c}}$ -- by using improved 
Schwinger-Dyson equations, or 
renormalization group equations (RGEs). A particularly transparent 
NJL-type framework, containing the essential features of the 
mentioned idea of condensation, is the Top-mode Standard Model 
(TSM) Lagrangian, known also as the BHL  
(Bardeen-Hill-Lindner) Lagrangian~\cite{bhl}.

In a recent work~\cite{cpv}, we studied the next-to-leading order 
(ntl) contributions in the 
$(1/N_{\mbox{\footnotesize c}})$-expansion
in the TSM by including quadratic fluctuations of
the composite Higgs $H=\langle \bar tt\rangle $ in the effective 
potential $V_{\mbox{\footnotesize eff}}$ . The existence of a 
non-trivial minimum in the 
effective potential led us to the conclusion that the cut-off 
$\Lambda $ is bound from above: 
$\Lambda \leq \Lambda _{crit} \approx 
4.7m_t^{\mbox{\footnotesize phys}}$ for 
$N_{\mbox{\footnotesize c}}=3$. 
In~\cite{cpv}, contributions related to components of the 
massive electroweak gauge bosons were not considered. QCD effects 
were included, but their impact was found to be small.
We considered the effective potential as a function
of a hard mass term $\lambda \sigma_0$ of the top quark, 
parametrized by the expectation value $\sigma_0$ of a composite
(initially auxiliary) scalar field $\sigma$.

In the present work, we continue the work of ref.~\cite{cpv}.
We show that the inclusion of the ``scalar'' longitudinal 
degrees of freedom of the electroweak gauge bosons $W$ 
and $Z$ (i.e., of the three composite Goldstones) at the ntl-level
does change the numerics substantially, but does not change the 
basic conclusion of the paper~\cite{cpv}. The cut-off remains in the 
region ${\cal {O}}(1\mbox{ TeV})$. As a matter of fact, the 
Goldstone contributions at the ntl-level tend to decrease the 
cut-off even further.

\section{The model and the effective potential}

In the Top-mode Standard Model (TSM) Lagrangian~\cite{bhl}, a 
truncated 
gauge-invariant 4-fermion interaction at a high energy scale 
$E \sim \Lambda$ is assumed to be responsible for the creation of 
a composite Higgs field $H=\langle \bar t t \rangle $ 
\begin{equation}
{\cal {L}}={\cal {L}}_{\mbox{\footnotesize kin}}^0
+G\left( \bar \Psi _{\mbox{\scriptsize L}}^{ia}
 t_{\mbox{\scriptsize R}a}\right) 
 \left( \bar t_{\mbox{\scriptsize R}}^b
 \Psi _{\mbox{\scriptsize L}b}^i\right) \qquad 
\mbox{for}\ E\sim \Lambda \ .  
\label{TSM}
\end{equation}
Here, $a$ and $b$ are the color and $i$ the isospin indices, 
$\Psi_{\mbox{\scriptsize L}}^T=(t_{\mbox{\scriptsize L}},
b_{\mbox{\scriptsize L}})$, and 
${\cal {L}}_{\mbox {\footnotesize kin}}^0$ contains the usual
gauge-invariant kinetic terms for fermions and gauge bosons. 
The Lagrangian (\ref{TSM}) leads to an effective framework for the 
minimal Standard Model. It can be rewritten in terms of an 
additional, as yet auxiliary, scalar isodoublet $\Phi $, by adding 
to it the following quadratic term~\footnote{
Addition of such a term changes the generating functional
only by an irrelevant source-independent factor~\cite{kugo}.} 
\[
{\cal {L}}_{\mbox {\footnotesize new}}=
{\cal {L}}_{\mbox{\footnotesize old}}-\left[ 
 M_0 \tilde \Phi ^{i\dagger }+\sqrt{G}
 \bar \Psi _{\mbox{\scriptsize L}}^{ia} 
 t_{\mbox{\scriptsize R}a}\right] \left[ 
 M_0{\tilde \Phi }^i+\sqrt{G}\bar 
 t_{\mbox{\scriptsize R}}^b{\Psi }_{\mbox{\scriptsize L} b}^i
 \right] \ ,
\]
\begin{equation}
\label{defPhi}\mbox{where:}\quad 
\tilde \Phi =i\tau _2\Phi ^{*}\ ,\qquad
\Phi =\frac 1{\sqrt{2}}\left( 
\begin{array}{c}
\sqrt{2}{\cal {G}}^{+} \\ {\cal H}+i{\cal {G}}^{(0)}
\end{array}
\right) \ ,\qquad 
{\cal {G}}^{\pm }= \frac{1}{\sqrt{2}}
 ({\cal {G}}^{(1)} \pm i {\cal {G}}^{(2)}) \ .
\end{equation}

The resulting Lagrangian reads 
\begin{eqnarray}
{\cal {L}} &=&i\bar \Psi ^a\partial \llap / \Psi _a-
\frac{M_0\sqrt{G}}{\sqrt{2}}\left[ 
{\cal H}\bar t^at_a-i{\cal {G}}^{(0)}\bar t^a\gamma _5t_a\right] +
\frac{M_0\sqrt{G}}2\left[ {\cal {G}}^{+}
\bar t^a(1-\gamma _5)b_a
+{\cal {G}}^{-}\bar b^a(1+\gamma _5)t_a\right]   
\nonumber \\
&&\ -\frac 12M_0^2\left( {\cal H}^2+{\cal {G}}^{(0)2}
+2{\cal {G}}^{-}{\cal {G}}^{+}\right) \ ,  
\label{TSM1}
\end{eqnarray}
where ${\cal H}$, ${\cal G}^{(0)}$, ${\cal G}^{(1)}$ and 
${\cal G}^{(2)}$ are the Higgs and the three real Goldstone 
components of the auxiliary complex isodoublet field $\Phi $, 
and $M_0$ is an unspecified bare mass term for $\Phi $ (at 
$E\sim \Lambda $)~\footnote{
The physical results will be independent of the value of $M_0^2$.}. 
These fields will eventually become the physical Higgs and the 
``scalar'' longitudinal components of the massive electroweak bosons 
through quantum effects. We ignore in (\ref{TSM1}) the transverse 
components of $W^{\pm }$ and $Z$ and all the lighter quarks which 
we assume to be and remain massless. It can be shown that the 
massless Goldstones discussed here correspond to the Goldstone 
degrees of freedom of $W^{\pm }$ and $Z$ in the Landau gauge 
($\xi \to \infty $); incidentally, in this gauge, 
the ghosts do not couple to the scalar degrees of freedom and 
therefore they (the ghosts) do not contribute to the effective 
potential~\cite{weinberg}.

The effective potential $V_{\mbox{\footnotesize eff}}(H_0)$ 
of the Higgs field ${\cal H}$ can then be calculated in 
Euclidean space by means of the following formula 
\begin{eqnarray}
\lefteqn{
\exp \left[ -\Omega V_{\mbox{\footnotesize eff}}(H_0)\right]
=const \times 
\int \prod_{j=0}^2 \left[ {\cal{D}}{\cal{G}}^{(j)}
\delta \left( \int d^4\bar y {\cal{G}}^{(j)}(\bar y)\right) 
\right] \times }  
\nonumber \\
&&
\times \int {\cal {D}}{\cal {H}}
\delta \left( \int d^4\bar y{\cal {H}} (\bar y)-\Omega H_0\right) 
\int {\cal {D}}\bar \Psi {\cal {D}} \Psi 
\exp \left[ +\int {\cal {L}}d^4\bar x\right] \ ,  
\label{pathint1}
\end{eqnarray}
where we set $\hbar = 1$.
The bars over space-time components, derivatives and momenta from 
now on denote Euclidean quantities. $\Omega $ is the 4-dimensional 
volume (formally infinite). We note that the effective potential is 
the energy density of the ground state when the order parameters 
$H_0=\langle {\cal H}\rangle $ and 
$\langle {\cal {G}}^{(j)}\rangle =0$ ($j=0,1,2$) are kept fixed. 
Next, we integrate out the quark degrees of freedom, and expand the 
resulting expression in powers of $h(\bar x)={\cal H}(\bar x)-H_0$, 
${\cal {G}}^{(j)}(\bar x)$ ($j=0,1,2$), including up to quadratic 
fluctuations. We thus obtain 
\begin{eqnarray}
\lefteqn{ \Omega V_{\mbox{\footnotesize eff}}(H_0)=
 \frac{1}{2} \Omega M_0^2 H_0^2  -Tr\ln \hat B_0
 - \ln \int_{-\infty }^\infty
\prod_{j=0}^2dJ_j\int_{-\infty }^\infty dJ_h
\int \prod_{j=0}^2 {\cal{D}}{\cal{G}}^{(j)} 
\int {\cal{D}}h }  
\nonumber \\
&& \exp {\Big \{}-\frac{M_0^2}2
 \int d^4\bar x\left[ h^2+{\cal {G}}^{(0)2}+
  2{\cal {G}}^{+}{\cal {G}}^{-}\right] 
 -\frac 12Tr\left( \hat B_0^{-1}\delta\hat B\right)^2
 -i\sum_{j=0}^2J_j\int d^4\bar x{\cal {G}}^{(j)}
 -iJ_h\int d^4 \bar xh
 {\Big \}} \ ,  
\label{pathint2}
\end{eqnarray}
where the integrals over $J_j$ ($j=0,1,2$) and $J_h$ represent the
corresponding $\delta $-functions in (\ref{pathint1}), and the
translationally invariant operator $\hat B_0$ as well as the scalar 
fluctuation operator $\delta \hat B$ can be written in the 
$\bar x$-basis as 
\begin{eqnarray}
\langle \bar x^{\prime };a|\hat B_0|\bar x;b\rangle  &=&\delta _{ab}
\left[ 
\begin{array}{cc}
\left( i{{\bar \partial }\llap /}+g_0 H_0/\sqrt{2}\right)  & 0 \\ 
0 & i{{\bar \partial }\llap /}
\end{array}
\right] \delta \left( \bar x-\bar x^{\prime }\right) \ ,  
\nonumber \\
\langle \bar x^{\prime };a|\delta \hat B|\bar x;b\rangle  &=&
\delta _{ab} \frac{g_0}{\sqrt{2}}\left[ 
\begin{array}{cc}
\left( h-i\gamma _5{\cal G}^{(0)}\right)  & -{\cal {G}}^{+}
(1-\gamma _5)/\sqrt{2} \\ 
-{\cal {G}}^{-}(1+\gamma _5)/\sqrt{2} & 0
\end{array}
\right] \delta \left( \bar x-\bar x^{\prime }\right) \ .  
\label{BdelB}
\end{eqnarray}
In (\ref{BdelB}), $a$ and $b$ are color indices, $g_0=M_0\sqrt{G}$, 
and the $2\times 2$ matrices are in isospin space. The first two 
terms on the r.h.s.~of (\ref{pathint2}) represent the 
leading-$N_{\mbox{\footnotesize c}}$ 
contribution $V_{\mbox{\footnotesize eff}}^{(0)}$ to the effective 
potential, while the exponential terms related to 
the quadratic fluctuations of the scalar fields lead to the 
full next-to-leading (ntl) contribution 
$V_{\mbox{\footnotesize eff}}^{(1)}$. The path integrals 
corresponding to these terms are of the Gaussian type and can be 
explicitly evaluated. Proceeding in close analogy to 
\cite{cpv}~\footnote{
In~\cite{cpv} we used for the ntl-contribution the notation
$V_{\mbox{\scriptsize eff}}^{\mbox{\scriptsize (ntl)}}$
(here: $V_{\mbox{\scriptsize eff}}^{(1)}$), and for the 
leading-$N_{\mbox{\scriptsize c}}$ (i.e., $0+1$-loop) contribution
the notation 
$V_{\mbox{\scriptsize eff}}^{(0)}+V_{\mbox{\scriptsize eff}}^{(1)}$
(here: $V_{\mbox{\scriptsize eff}}^{(0)}$).}, 
we obtain the following: 
\begin{equation}
V_{\mbox{\footnotesize eff}}=
 V_{\mbox{\footnotesize eff}}^{(0)}
 +V_{\mbox{\footnotesize eff}}^{(1)} 
 + {\cal {O}}(1/N^2_{\mbox{\footnotesize c}}) ,  
\label{V01}
\end{equation}
where the leading-$N_{\mbox{\footnotesize c}}$ contribution is 
\begin{equation}
V_{\mbox{\footnotesize eff}}^{(0)}\left( \lambda ^2\sigma _0^2\right) 
= \sigma _0^2
-\frac{N_{\mbox{\footnotesize c}}}{8\pi ^2} \int_0^{\Lambda^2_f} 
d{\bar k}^2 {\bar k}^2
\ln \left[ 1+\frac{\lambda^2\sigma _0^2}{{\bar k}^2}\right] \ ,   
\label{V0}
\end{equation}
and the ntl-term is 
\begin{equation}
V_{\mbox{\footnotesize eff}}^{(1)}\left( \lambda ^2\sigma _0^2\right) 
= \left[ 
\frac 1{2\Omega} Tr\ln \left( 
 {\hat A}_{\mbox{\footnotesize n}}\right)_{11}+
\frac 1{2\Omega }Tr\ln \left( 
 {\hat A}_{\mbox{\footnotesize n}}\right) _{22}+
\frac 2{2\Omega }Tr\ln \left( 
 {\hat A}_{\mbox{\footnotesize ch}}\right) \right] \ .
\label{V1}
\end{equation}
We denoted 
\begin{equation}
\lambda =\sqrt{G}\ ,\quad \sigma_0=M_0 H_0/\sqrt{2}\ ,  
\label{not}
\end{equation}
and ${\hat A}_{\mbox{\footnotesize n}}$ and 
${\hat A}_{\mbox{\footnotesize ch}}$ are the kernels of the Gaussian 
path integrals corresponding to the contributions of the neutral 
scalars ${\cal {H}}$ , ${\cal {G}}^{(0)}$, and charged scalars 
${\cal {G}}^{\pm }$, respectively: 
\begin{equation}
\langle {\bar x}^{\prime }|
{\hat A}_{\mbox{\footnotesize n,ch}}|\bar x \rangle =
\frac 1{(2\pi)^4} \int d^4{\bar p}
\exp \left[ i{\bar p}\left( {\bar x}-{\bar x}^{\prime}\right) \right] 
{\tilde A}_{\mbox{\footnotesize n,ch}}\left( {\bar p}^2\right) \ .  
\label{AFour}
\end{equation}
The Fourier transforms of these kernels are 
\begin{eqnarray}
\tilde A_{\mbox{\footnotesize n,ch}}\left( {\bar p}^2\right)  &=&
\int d^4{\bar x} \exp \left[ -i{\bar p}\cdot {\bar x}\right] 
\langle 0 | {\hat A}_{\mbox{\footnotesize n,ch}} | {\bar x} \rangle =
2 \left[ {\hat 1}-
2\lambda^2 N_{\mbox{\footnotesize c}} 
{\cal K}_{\mbox{\footnotesize n,ch}} 
\left( {\bar p}^2;\lambda^2 \sigma_0^2 \right) \right] \ ,  
\nonumber \\
{\cal K}_{\mbox{\footnotesize n}} &=&
\frac{1}{4} \int_{{\bar k}^2 \leq \Lambda_{\mbox{\footnotesize f}}^2}
\frac{d^4{\bar k}}{(2\pi )^4} \left[ 
\begin{array}{cc}
tr_{\mbox{\scriptsize f}} \left( 
 \frac i{({{\bar k}\llap /}-\lambda \sigma _0)}
 \frac i{({{\bar p} \llap /}+{{\bar k}\llap /}-\lambda \sigma _0)}
 \right)  
& 0 \\ 
0 & 
tr_{\mbox{\scriptsize f}}
\left( \frac i{({{\bar k}\llap /}-\lambda \sigma _0)}
 \left( -i\gamma_5\right) 
 \frac i{({{\bar p}\llap /}+{{\bar k}\llap /}-\lambda \sigma_0)}
\left( -i\gamma _5\right) \right) 
\end{array}
\right] \ ,   
\nonumber \\
{\cal K}_{\mbox{\footnotesize ch}} &=&
\frac{1}{2} \int_{{\bar k}^2 \leq \Lambda_{\mbox{\footnotesize f}}^2}
\frac{d^4{\bar k}}{(2\pi )^4}
tr_{\mbox{\scriptsize f}} 
\left[ \frac i{({{\bar k}\llap /}-\lambda \sigma_0)}
 \left( \frac{1-\gamma_5}{2} \right) 
 \frac i{({{\bar p}\llap /}+{{\bar k}\llap /})}
\left( \frac{1+\gamma_5}{2} \right) \right] \ .  
\label{Ks}
\end{eqnarray}
Here, $tr_{\mbox{\scriptsize f}}$ means tracing over the 
$4\times 4$ spinor matrices. 
The integrals in (\ref{Ks}), as well as the one in (\ref{V0}),
are regularized by means of a simple spherical cut-off 
$\Lambda_{\mbox{\footnotesize f}}$ 
for the fermionic (top quark) momenta $|{\bar k}|$. We note that 
$({\cal K}_{\mbox{\footnotesize n}})_{11}$ and 
$({\cal K}_{\mbox{\footnotesize n}})_{22}$ are truncated 2-point 
Green functions corresponding to a $(t{\bar t})$-loop carrying two 
external Higgs legs, and two neutral Goldstone legs (with momentum 
${\bar p}$), respectively. Analogously, 
${\cal K}_{\mbox{\footnotesize ch}}$ corresponds 
to a $(b{\bar t})$-loop with two external legs of the charged 
Goldstones. The tracing over the colors led to factors 
$N_{\mbox{\footnotesize c}}$ ($=3$) in 
$V_{\mbox{\footnotesize eff}}^{(0)}$ and in front of 
${\cal {K}}_{\mbox{\footnotesize n,ch}}$ 
in $V_{\mbox{\footnotesize eff}}^{(1)}$. 
The tracing in (\ref{V1}) is over spinor space 
and the momentum basis, involving a second integral over the 
bosonic momenta ${\bar p}$ (cf.~also ref.~\cite{cpv}). We introduce 
for these momenta a second spherical cut-off: 
${\bar p}^2\leq \Lambda_{\mbox{\footnotesize b}}^2$, 
where subscript ``b'' stands for
``bosonic'' (note: ${\Lambda}_{\mbox{\footnotesize b}}\sim 
{\Lambda}_{\mbox{\footnotesize f}}$). We then 
rescale all the momenta 
$\{{\bar k}^2,{\bar p}^2\}\to 
\Lambda_{\mbox{\footnotesize f}}^2\{{\bar k}^2,{\bar p}^2\}$, 
and introduce the following dimensionless quantities: 
\[
{\varepsilon}^2=
\lambda^2\frac{\sigma_0^2}{\Lambda_{\mbox{\footnotesize f}}^2}=
\frac{GM_0^2}{2\Lambda_{\mbox{\footnotesize f}}^2}H_0^2\ , \qquad 
a=\frac{(GN_{\mbox{\footnotesize c}}
\Lambda_{\mbox{\footnotesize f}}^2)}{(8\pi^2)}\ ,
\]
\begin{equation}
\Xi_{\mbox{\footnotesize eff}}=
8\pi ^2 V_{\mbox{\footnotesize eff}}/
(N_{\mbox{\footnotesize c}}\Lambda_{\mbox{\footnotesize f}}^4)=
\Xi ^{(0)}+\frac 1{N_{\mbox{\footnotesize c}}}\Xi^{(1)}
+ {\cal {O}}(\frac{1}{N^2_{\mbox{\footnotesize c}}}) \ .  
\label{defka}
\end{equation}
The resulting expressions for the 
leading-$N_{\mbox{\footnotesize c}}$ term $\Xi ^{(0)}$ 
and the ntl-term $\Xi ^{(1)}$ are
\begin{eqnarray}
\Xi ^{(0)}\left( {\varepsilon}^2;a\right) =
\frac{{\varepsilon}^2}a
 -\int_0^1d{\bar k}^2{\bar k}^2
\ln \left( 1+\frac{{\varepsilon}^2}{{\bar k}^2}\right) \ ,  
\label{Xi0}
\end{eqnarray}
\begin{eqnarray}
\lefteqn{
\Xi^{(1)}\left( {\varepsilon}^2;
{\Lambda }_{\mbox{\footnotesize b}}^2/
{\Lambda }_{\mbox{\footnotesize f}}^2;a \right) =
{\Big\{} 
\frac{1}{4} \int_0^{{\Lambda }_{\mbox{\scriptsize b}}^2/
{\Lambda }_{\mbox{\scriptsize f}}^2} 
 d\bar p^2 \bar p^2 \ln \left[ 
  1-a {\cal J}_{H} \left( {\bar p}^2; {\varepsilon}^2\right) \right]  
}  
\nonumber \\
&&
+\frac{1}{4} \int_0^{{\Lambda }_{\mbox{\scriptsize b}}^2/
{\Lambda}_{\mbox{\scriptsize f}}^2} 
 d{\bar p}^2 {\bar p}^2 \ln \left[ 
  1-a{\cal J}_{Gn}\left( {\bar p}^2;{\varepsilon}^2\right) \right]
+\frac{2}{4} \int_0^{{\Lambda }_{\mbox{\scriptsize b}}^2/
{\Lambda}_{\mbox{\scriptsize f}}^2} 
 d{\bar p}^2 {\bar p}^2 \ln \left[ 
  1-a{\cal J}_{Gch}\left( {\bar p}^2;{\varepsilon}^2\right) \right] 
{\Big\}} \ .  
\label{Xi1}
\end{eqnarray}
Subscripts $H$, $Gn$ and $Gch$ correspond to  contributions from 
the Higgs, neutral Goldstone and  charged Goldstone degrees of 
freedom, respectively. The dimensionless 2-point Green functions 
are defined as 
\begin{eqnarray}
{\cal J}_H\left( {\bar p}^2;{\varepsilon}^2\right)  &=&
\frac{16\pi ^2}{{\Lambda}_{\mbox{\footnotesize f}}^2}
{\cal K}_{\mbox{\footnotesize n}}
\left( {\Lambda}_{\mbox{\footnotesize f}}^2{\bar p}^2;
 \lambda ^2\sigma_0^2\right)_{11}=
\frac 1{\pi ^2}\int_{{\bar k}^2\leq 1}d^4{\bar k}
\frac{\left[ {\bar k}\cdot ({\bar p}+{\bar k})-{\varepsilon}^2
 \right] }
{\left( {\bar k}^2+{\varepsilon}^2\right) 
\left[ ({\bar p}+{\bar k})^2+{\varepsilon}^2\right] }\ ,  
\nonumber \\
{\cal J}_{Gn}\left( {\bar p}^2;{\varepsilon}^2\right)  &=&
\frac{16\pi ^2}{{\Lambda}_{\mbox{\footnotesize f}}^2}
{\cal K}_{\mbox{\footnotesize n}}
\left( {\Lambda}_{\mbox{\footnotesize f}}^2{\bar p}^2;
 \lambda ^2\sigma_0^2\right)_{22}=
\frac 1{\pi ^2}\int_{{\bar k}^2\leq 1}d^4{\bar k}
\frac{\left[ {\bar k}\cdot ({\bar p}+{\bar k})+{\varepsilon}^2
 \right] }
{\left( {\bar k}^2+{\varepsilon}^2\right) 
\left[ ({\bar p}+{\bar k})^2+{\varepsilon}^2\right] }\ ,  
\nonumber \\
{\cal J}_{Gch}\left( {\bar p}^2;{\varepsilon}^2\right)  &=&
\frac{16\pi ^2}{{\Lambda}_{\mbox{\footnotesize f}}^2}
{\cal K}_{\mbox{\footnotesize ch}}
\left( {\Lambda}_{\mbox{\footnotesize f}}^2{\bar p}^2;
 \lambda ^2\sigma_0^2\right)_{11}=
\frac 1{\pi^2}\int_{{\bar k}^2\leq 1}d^4{\bar k}
\frac{{\bar k}\cdot ({\bar p}+{\bar k})}
{\left( {\bar k}^2+{\varepsilon}^2\right) ({\bar p}+{\bar k})^2}\ .  
\label{Js}
\end{eqnarray}
The expressions for $\Xi^{(0)}$ and $\Xi^{(1)}$ can also be
rederived diagrammatically by summing up terms corresponding
to the 1-PI Green functions depicted in Figs.~1 and 2
(cf.~also~\cite{cpv}). 

The minimization of the leading-$N_{\mbox{\footnotesize c}}$ 
part of $\Xi_{\mbox{\footnotesize eff}} $ leads to the familiar 
leading-$N_{\mbox{\footnotesize c}}$ gap equation connecting the 
cut-off ${\Lambda}_{\mbox{\footnotesize f}}$, the 4-fermion 
coupling strength $G$ and the 
leading-$N_{\mbox{\footnotesize c}}$ 
approximation $m_t^{(0)}$ to the mass of the top quark 
\[
\frac{\partial \Xi ^{(0)}\left( {\varepsilon}^2;a\right) }
{\partial {\varepsilon}^2}{\Big |}_{{\varepsilon}^2=
{\varepsilon}_0^2}=0\ ,\quad
\Rightarrow 
\]
\begin{equation}
\Rightarrow \qquad a\quad 
\left( =\frac{GN_{\mbox{\footnotesize c}}
{\Lambda}_{\mbox{\footnotesize f}}^2}{8\pi ^2}\right)
=\left[ 
1-{\varepsilon}_0^2\ln \left( {\varepsilon}_0^{-2}+1\right)
\right]^{-1} \quad 
\left( ={\cal {O}}(1)\right) \ ,\quad \mbox{where: }
{\varepsilon}_0^2=
\left( m_t^{(0)}/{\Lambda}_{\mbox{\footnotesize f}}\right)^2 \ .  
\label{gaplead}
\end{equation}
This condition shows that the parameter $a(>1)$ is a number of 
${\cal {O}}(1)$, and should be regarded as a quantity 
${\cal {O}}(N_{\mbox{\footnotesize c}}^0)$ in the 
$(1/N_{\mbox{\footnotesize c}})$-expansion, as already
done in eqs.~(\ref{defka})-(\ref{Xi1}). Next-to-leading order
information connecting the bare mass 
$m_t({\Lambda}_{\mbox{\footnotesize f}})$, the cut-off 
${\Lambda}_{\mbox{\footnotesize f}}$ and the 4-fermion 
coupling strength $G$, can be obtained by consistently
minimizing $\Xi_{\mbox{\footnotesize eff}}$ at each order
in $(1/N_{\mbox{\footnotesize c}})$-expansion. 
\begin{equation}
\frac{\partial \Xi _{\mbox{\footnotesize eff}}
\left( {\varepsilon}^2;a \right) }
{\partial {\varepsilon}^2}
{\Big |}_{{\varepsilon}^2=
{\varepsilon}^2_{\mbox{\scriptsize gap}}} = 0 \ ,
\label{gap}
\end{equation}
\begin{equation}
\mbox{where: } \quad 
{\varepsilon}^2_{\mbox{\scriptsize gap}}
=\frac{m_t^2({\Lambda}_{\mbox{\footnotesize f}})}
{{\Lambda}_{\mbox{\footnotesize f}}^2}=
{\varepsilon}_0^2+\frac{1}{N_{\mbox{\footnotesize c}}}
\kappa _{1\mbox{\footnotesize g}}+
{\cal {O}}(\frac{1}{N_{\mbox{\footnotesize c}}^2}) \ .  
\label{epsexpan}
\end{equation}
Inserting (\ref{epsexpan}) into (\ref{gap}), taking into account
(\ref{defka}) for $\Xi_{\mbox{\footnotesize eff}}$ and demanding
that the coefficients at each power of
$(1/N_{\mbox{\footnotesize c}})$ are zero, we obtain the following 
relations: 
\[
\frac{\partial \Xi ^{(0)}}{\partial {\varepsilon}^2}
{\Big |}_{{\varepsilon}^2={\varepsilon}_0^2}=0 \ ,
\]
\begin{equation}
\kappa_{1\mbox{\footnotesize g}} \frac{\partial^2\Xi^{(0)}}
{\partial ({\varepsilon}^2)^2}
{\Big |}_{{\varepsilon}^2={\varepsilon}_0^2}+
\frac{\partial \Xi ^{(1)}}{\partial {\varepsilon}^2}
{\Big |}_{{\varepsilon}^2={\varepsilon}_0^2} = 0 \ .
\label{gap1}
\end{equation}
The ntl-gap equation (\ref{gap1}) determines the change of 
the ratio 
${\varepsilon}^2_{\mbox{\scriptsize gap}}=
m_t^2({\Lambda}_{\mbox{\footnotesize f}})/
{\Lambda}^2_{\mbox{\footnotesize f}}$ 
due to ntl-effects 
\begin{equation}
\delta ({\varepsilon}^2)^{\mbox{\scriptsize (ntl)}}
_{\mbox{\scriptsize gap}} =
\frac{\kappa _{1\mbox{\footnotesize g}}}
{N_{\mbox{\footnotesize c}}}=
-\left[ 
\frac{\partial \Xi ^{(1)}}{\partial {\varepsilon}^2}
{\Big |}_{{\varepsilon}^2={\varepsilon}_0^2}
\right] {\Bigg /}
\left[ N_{\mbox{\footnotesize c}}
  \frac{\partial ^2\Xi ^{(0)}}{\partial ({\varepsilon}^2)^2}
{\Big |}_{{\varepsilon}^2={\varepsilon}_0^2}\right] \ .  
\label{deps2gap}
\end{equation}

Next, we turn to mass renormalization corrections: 
$m_t({\Lambda}_{\mbox{\footnotesize f}})$ 
$\mapsto (m_t)_{\mbox{\scriptsize  ren.}}$.
It is straightforward to check that there are no
leading-$N_{\mbox{\footnotesize c}}$ contributions to these
corrections, so that only the 1-PI diagrams shown in Fig.~3
must be taken into account (cf.~also~\cite{cpv}). 
\begin{equation}
\delta ({\varepsilon}^2)_{\mbox{\scriptsize ren.}}=
\frac{(m_t^2)_{\mbox{\scriptsize ren.}}}
 {{\Lambda}_{\mbox{\footnotesize f}}^2}-
 \frac{m_t^2({\Lambda}_{\mbox{\footnotesize f}})}
 {{\Lambda}_{\mbox{\footnotesize f}}^2}=
\frac{1}{N_{\mbox{\footnotesize c}}}
 \kappa _{1\mbox{\footnotesize r}}
 +{\cal {O}}(\frac{1}{N_{\mbox{\footnotesize c}}^2}) \ .
\label{deps1ren}
\end{equation}
At the ntl-level there are three separate contributions, coming 
from the Higgs, neutral Goldstone and the charged Goldstone,
respectively (cf.~Fig.~3)
\begin{equation}
\delta ( {\varepsilon}^2 )_{\mbox{\scriptsize ren.}}
^{\mbox{\scriptsize (ntl)}} =
\frac{1}{N_{\mbox{\footnotesize c}}} 
\kappa_{1\mbox{\footnotesize r}} =
\frac{1}{N_{\mbox{\footnotesize c}}} \left(
\kappa _{1\mbox{\footnotesize r}}^{(H)}
+\kappa _{1\mbox{\footnotesize r}}^{(Gn)}
+\kappa _{1\mbox{\footnotesize r}}^{(Gch)}
\right) \ .
\label{kappar}
\end{equation}
Calculations and summations of the diagrams of Fig.~3 in
Euclidean space yield 
\begin{equation}
\kappa _{1\mbox{\footnotesize r}}^{(H)}=-\frac{a}{4}
\int_0^{{\Lambda }_{\mbox{\scriptsize b}}^2/
{\Lambda}_{\mbox{\scriptsize f}}^2}
\frac{d{\bar p}^2}
{\left[ 1-a{\cal J}_H\left( {\bar p}^2;{\varepsilon}_0^2\right)
\right] }
\left[ \left( \sqrt{ {\bar p}^2 ( {\bar p}^2+4{\varepsilon}_0^2 ) }
-{\bar p}^2 \right) 
\left( 2+\frac{{\bar p}^2}{2{\varepsilon}_0^2}\right) 
- {\bar p}^2 \right] \ ,  
\label{kapHr}
\end{equation}
\begin{equation}
\kappa _{1\mbox{\footnotesize r}}^{(Gn)}=
+\frac a4\int_0^{{\Lambda }_{\mbox{\scriptsize b}}^2/
{\Lambda}_{\mbox{\scriptsize f}}^2}
\frac{d{\bar p}^2}
{\left[ 1-a{\cal J}_{Gn}\left( {\bar p}^2;{\varepsilon}_0^2\right) 
\right] }
\frac{1}{2{\varepsilon}_0^2}
{\bar p}^2
\left[ {\bar p}^2+2{\varepsilon}_0^2
-\sqrt{ {\bar p}^2 ( {\bar p}^2+4{\varepsilon}_0^2 ) }
\right] \ ,  
\label{kapGnr}
\end{equation}
\begin{eqnarray}
\kappa _{1\mbox{\footnotesize r}}^{(Gch)} &=&
+\frac{a}{4}
{\Big \{}
\int_0^{-{\varepsilon}_0^2} 
\frac{d{\bar p}^2{\bar p}^2 
 \left[ 2+{\bar p}^2/{\varepsilon}_0^2 \right] }
{\left[ 1-a{\cal J}_{Gch}\left( {\bar p}^2;{\varepsilon}_0^2
\right) \right] }
-{\varepsilon}_0^2 
 \int_{-{\varepsilon}_0^2}^{{\Lambda }
_{\mbox{\scriptsize b}}^2/{\Lambda}_{\mbox{\scriptsize f}}^2}
d{\bar p}^2 \left[ 
\frac{1}{\left( 
1-a{\cal J}_{Gch}\left( {\bar p}^2;{\varepsilon}_0^2\right) 
\right) }
-\frac{2}{{\bar p}^2 a
\ln \left( 1+1/{\varepsilon}_0^2\right) }
\right]   
\nonumber \\
&& 
-\frac{2{\varepsilon}_0^2}
{a\ln \left( 1+1/{\varepsilon}_0^2\right) }
\left[ -\ln {\varepsilon}_0^2+
 \ln \left( {\Lambda }_{\mbox{\footnotesize b}}^2/
{\Lambda}_{\mbox{\footnotesize f}}^2\right) \right] 
{\Big \}} \ .  
\label{kapGchr}
\end{eqnarray}
The expressions above were obtained by summing up the corresponding 
Green functions of Fig.~3, assuming first a (normalized) Euclidean
momentum ${\bar q}^2>0$ for the external top quark line. Then the 
analytic continuation to the (approximate) on-shell values 
${\bar q}^2=-q^2=-m_t^{(0)2}/{\Lambda}^2_{\mbox{\footnotesize f}} 
(=-{\varepsilon}^2_0)$ had to be performed. 
In the case of $\kappa^{(H)}_{1\mbox{\footnotesize r}}$ and 
$\kappa^{(Gn)}_{1\mbox{\footnotesize r}}$,
it turned out that this continuation is equivalent to the simple
substitution in the Euclidean integrands: 
${\bar q}^2 \mapsto -{\varepsilon}^2_0$. The contribution
$\kappa^{(Gch)}_{1\mbox{\footnotesize r}}$ 
of the charged Goldstones leading to 
(\ref{kapGchr}) is somewhat more complicated due to the fact that
the massless Goldstone pole at ${\bar p}^2 = 0$ generates a 
logarithmic branch cut in 
$\kappa^{(Gch)}_{1\mbox{\footnotesize r}}({\bar q}^2)$ 
at the threshold value ${\bar q}^2 = 0$. The analytic continuation
follows then the usual prescription:
$\ln {\bar q}^2 \mapsto \ln ( -q^2- i \epsilon ) \mapsto
\ln q^2 - i \pi^2$, for $q^2 > 0 $. The real part of this term
was written in (\ref{kapGchr}) as a separate 
$\ln {\varepsilon}^2_0$-term. Therefore, none of the remaining
integrals in (\ref{kapGchr}) is singular.

A few comments are in order here. Eq.~(\ref{deps2gap}) shows that
the magnitude of the ntl-corrections 
$\delta ( {\varepsilon}^2 )_{\mbox{\scriptsize gap}}
^{\mbox{\scriptsize (ntl)}}$ depends strongly on 
${\varepsilon}_0^2$, i.e., the solution of the 
leading-$N_{\mbox{\footnotesize c}}$ gap equation. This is
consistent, since all our calculations were carried out in the
``$(1/N_{\mbox{\footnotesize c}})$-perturbative'' manner.
The integrals involved in (\ref{deps2gap}) contain no singularities.
On the other hand, if we were to relax the 
large-$N_{\mbox{\footnotesize c}}$ expansion (\ref{epsexpan})
and solve (\ref{gap}) (with: $\Xi_{\mbox{\footnotesize eff}} =
\Xi^{(0)}+\Xi^{(1)}/N_{\mbox{\footnotesize c}}$) without assuming
(\ref{epsexpan}), we would encounter singularities in the integrals
over ${\bar p}^2$, suggesting that such an approach does not
guarantee the masslessness of the Goldstones. These singularities
would correspond to the appearance of small nonzero squares of
masses for the Goldstones, and they would cancel away only
when higher order terms 
${\cal {O}}(1/N^2_{\mbox{\footnotesize c}})$ were included in
$\Xi_{\mbox{\footnotesize eff}}$. Analogous considerations apply
also to $\delta ( {\varepsilon}^2 )_{\mbox{\scriptsize ren.}}$.

We have numerically calculated the ntl-changes (\ref{deps2gap}) 
and (\ref{kappar}), based on the integrals (\ref{Xi0})-(\ref{Xi1}) 
and (\ref{kapHr})-(\ref{kapGchr}). The integrals over the 
squares ${\bar p}^2$ of the normalized bosonic momenta were 
performed using the following explicit expressions for the 
normalized 2-point Green functions (\ref{Js}) 
\begin{eqnarray}
{\cal {J}}_{Gch}\left( {\bar p}^2;{\varepsilon}^2\right)  &=&
\left[ 1-\frac34{\bar p}^2-\frac 12{\varepsilon}^2
+\frac 1{2{\bar p}^2} 
\left( {\bar p}^2+{\varepsilon}^2 \right)^2
\ln ({\bar p}^2+{\varepsilon}^2)
-\frac 12\left( 2{\varepsilon}^2+{\bar p}^2\right) 
\ln (1+{\varepsilon}^2)
-\frac{{\varepsilon}^4}{2{\bar p}^2}\ln {\varepsilon}^2
\right] \ ,  
\nonumber \\
{\cal {J}}_H\left( {\bar p}^2;{\varepsilon}^2\right)  &=&
{\cal {D}}\left( {\bar p}^2;{\varepsilon}^2 \right) 
 -\frac{({\bar p}^2+4{\varepsilon}^2)}4
{\cal {C}}\left( {\bar p}^2;{\varepsilon}^2\right) \ ,  
\nonumber \\
{\cal {J}}_{Gn}\left( {\bar p}^2;{\varepsilon}^2\right)  &=&
{\cal {D}}\left( {\bar p}^2;{\varepsilon}^2 \right) 
 -\frac{{\bar p}^2}4{\cal {C}}
\left( {\bar p}^2;{\varepsilon}^2\right) \ ,  
\label{Jexpl}
\end{eqnarray}
where we have defined: 
\begin{eqnarray}
{\cal {D}}\left( {\bar p}^2;{\varepsilon}^2\right)  &=&
\left[ \frac 34
 -\frac{{\varepsilon}^2}2\ln ({\varepsilon}^{-2}+1)
 +\frac 1{8{\bar p}^2}(1+{\varepsilon}^2)^2
 -\frac{{\bar p}^2}8
 -\frac 1{8{\bar p}^2}(1-{\bar p}^2+{\varepsilon}^2){\cal {B}}
 -\frac{{\varepsilon}^2}2
 \ln \left( \frac{a_3}{2{\varepsilon}^2}\right) 
\right] \ ,  
\nonumber \\
{\cal {C}}\left( {\bar p}^2;{\varepsilon}^2\right)  &=&
\left[ 1
 +\frac 1{{\bar p}^2}(1+{\varepsilon}^2-{\cal {B}})
 +(1-{\cal {A}})\ln ({\varepsilon}^{-2}+1)
 +{\cal {A}}\ln \left( \frac{a_1}{a_2} \right) 
 +\ln \left( \frac{a_3}{2{\varepsilon}^2}\right) 
\right] \ .  
\label{Jexpl2}
\end{eqnarray}
The parameters ${\cal {A}}$, ${\cal {B}}$ and $a_j$ ($j=1,2,3$) 
denote the expressions 
\[
{\cal {A}}=\sqrt{1+4\frac{{\varepsilon}^2}{{\bar p}^2}} \ ,
\quad 
{\cal {B}}=\sqrt{(1-{\bar p}^2+{\varepsilon}^2)^2
 +4{\bar p}^2{\varepsilon}^2} \ ,
\]
\begin{equation}
a_1=(\bar p^2+3{\varepsilon}^2-1+{\cal {A}}{\cal {B}}) \ ,
\ \ a_2=\bar p^2+3{\varepsilon}^2
 +(\bar p^2+{\varepsilon}^2){\cal {A}} \ ,
\ \ 
a_3=1-\bar p^2+{\varepsilon}^2+{\cal {B}}
\ .  
\label{Jexpl3}
\end{equation}
The partial derivatives 
$\partial {\cal {J}}_X / \partial {\varepsilon}^2$ 
($X=H$, $Gn$, $Gch$), needed for the calculation of the integrand 
of $\partial \Xi^{(1)} / \partial {\varepsilon}^2$ in the ntl-gap
equation (\ref{gap1})-(\ref{deps2gap}), are obtained directly
from (\ref{Jexpl}).

The input value for the integrations was the parameter 
$a=N_{\mbox{\footnotesize c}}G{\Lambda}_
{\mbox{\footnotesize f}}^2/8\pi ^2$ of 
(\ref{defka}) and (\ref{gaplead}), which is essentially a 
dimensionless measure of the strength of the original 
4-fermion coupling $G$ in (\ref{TSM}). 
We also had to choose a specific value of the ratio of the
cut-offs ${\Lambda }_{\mbox{\footnotesize b}}/
{\Lambda}_{\mbox{\footnotesize f}}$ ($={\cal {O}}(1)$). As suggested 
by the diagrams of Fig.~2, realistic choices in the present 
framework of simple spherical cut-offs have: 
${\bar p}_{\mbox{\footnotesize max}}^2
\leq {\bar k}_{\mbox{\footnotesize max}}^2$, which implies 
${\Lambda }_{\mbox{\footnotesize b}}/
{\Lambda}_{\mbox{\footnotesize f}} \stackrel{<}{\sim} 1$. 
We have made two choices: 
${\Lambda }_{\mbox{\footnotesize b}}/
{\Lambda}_{\mbox{\footnotesize f}}=1/\sqrt{2}$ ($\approx 0.707$), 
$0.5$. It turned out 
that the ntl-effects (\ref{deps2gap}) and (\ref{kappar}) 
decrease the ratio 
$(m_t^{\mbox{\scriptsize phys.}}/
{\Lambda}_{\mbox{\footnotesize f}})^2$ 
when compared to the leading-$N_{\mbox{\footnotesize c}}$ 
expression ${\varepsilon}_0^2$ of (\ref{gaplead}) (note: at the 
leading-$N_{\mbox{\footnotesize c}}$ level we have 
$m_t^{(0)}=m_t^{\mbox{\scriptsize phys.}}$, 
and at the ntl-level we have 
$m_t^{\mbox{\scriptsize  ren.}}=m_t^{\mbox{\scriptsize phys.}}$). 
Stability of the results requires that the ntl-changes of the ratio 
$m_t^{\mbox{\scriptsize phys.}}/{\Lambda}_{\mbox{\footnotesize f}}$ 
not be too large, so that the 
$(1/N_{\mbox{\footnotesize c}})$-expansion would have some 
qualitative predictive power. This implies that the parameter $a$, 
or equivalently the leading-$N_{\mbox{\footnotesize c}}$ ratio 
${\varepsilon}_0=m_t^{(0)}/{\Lambda}_{\mbox{\footnotesize f}}$ 
(cf.~(\ref{gaplead})), cannot decrease beyond a certain critical 
value, and that the resulting ratio 
$m_t^{\mbox{\scriptsize  ren.}}/{\Lambda}_{\mbox{\footnotesize f}}$ 
cannot be smaller than a critical value 
$(m_t^{\mbox{\scriptsize ren.}}/
{\Lambda}_{\mbox{\footnotesize f}})_{\mbox{\footnotesize crit.}}$ 
correspondingly.  
Consequently, the cut-off ${\Lambda}_{\mbox{\footnotesize f}}$ 
cannot exceed an upper bound 
$({\Lambda}_{\mbox{\footnotesize f}})_{\mbox{\footnotesize max}}$ 
(we took: 
$m_t^{\mbox{\scriptsize  ren.}}=$
$m_t^{\mbox{\scriptsize phys.}}=180\mbox{ GeV}$). 
Specifically, we demanded that the value of 
$m_t^{\mbox{\scriptsize phys.}}/{\Lambda}_{\mbox{\footnotesize f}}$ 
be diminished by the ntl-effects 
(\ref{deps2gap}) and (\ref{kappar}) not
more than by a factor of: $\sqrt{2}$, $2$, $3$, $4$. The 
resulting critical values of ratios and of cut-offs are given 
in Table 1 (columns 3-6), where, in addition, we included in the 
last four columns the results when only the Higgs effects (without 
Goldstones) were taken into account. Comparing the two sets of 
results, we conclude that the Goldstone degrees of freedom
change the numbers substantially. However, in both cases, we are 
led to the same qualitative conclusion: the cut-off 
${\Lambda}_{\mbox{\footnotesize f}}$ does not surpass 
${\cal {O}}(1\mbox{ TeV})$. For 
${\Lambda}_{\mbox{\footnotesize b}}/
{\Lambda}_{\mbox{\footnotesize f}} = 1$ the calculations show
that the negative ntl-contribution
$\delta ( {\varepsilon}^2 )^{\mbox{\scriptsize (ntl)}} $
(=$\delta ( {\varepsilon}^2 )_{\mbox{\scriptsize gap}}
^{\mbox{\scriptsize (ntl)}} +
\delta ( {\varepsilon}^2 )_{\mbox{\scriptsize ren.}}
^{\mbox{\scriptsize (ntl)}}$) is under the inclusion of the
Goldstone contributions always stronger than the
leading-$N_{\mbox{\footnotesize c}}$ one:
$|\delta ( {\varepsilon}^2 )^{\mbox{\scriptsize (ntl)}}|
> {\varepsilon}_0^2$.

Looking more closely upon the contributions of the various degrees 
of freedom to the ``gap'' ntl-shift 
$\delta ({\varepsilon}^2)_{\mbox{\scriptsize gap}}$ 
of (\ref{deps2gap}) and to the mass renormalization ntl-shift 
$\delta ({\varepsilon}^2)_{\mbox{\scriptsize  ren.}}$ of 
(\ref{kappar}), for the 
cases displayed in Table 1, the following picture emerges: 
the Higgs and each one of the three Goldstone degrees of freedom 
contribute comparable negative values to 
$\delta ({\varepsilon}^2)_{\mbox{\scriptsize gap}}$; 
the Higgs and the charged Goldstone degrees of freedom
contribute each a negative value and the 
neutral Goldstone a weaker positive value
to $\delta ({\varepsilon}^2)_{\mbox{\scriptsize  ren.}}$, 
leading thus to a negative
$\delta ({\varepsilon}^2)_{\mbox{\scriptsize  ren.}}$.
Consequently, both 
$\delta ({\varepsilon}^2)_{\mbox{\scriptsize gap}}$ 
and $\delta ({\varepsilon}^2)_{\mbox{\scriptsize ren.}}$ 
are negative,
and $|\delta ({\varepsilon}^2)_{\mbox{\scriptsize  gap}}|$ is
larger than
$|\delta ({\varepsilon}^2)_{\mbox{\scriptsize  ren.}}|$, usually
by more than a factor of $2$
($\delta ({\varepsilon}^2)_{\mbox{\scriptsize gap}} \approx -0.3$, 
$-0.15$, $-0.05$, for 
${\Lambda}_{\mbox{\footnotesize b}}/
{\Lambda}_{\mbox{\footnotesize f}} = 1$, 
$0.707$, $0.5$, respectively).
It turns out that
$|\delta ({\varepsilon}^2)_{\mbox{\scriptsize  ren.}}|/
{\varepsilon}_0^2 \stackrel{<}{\sim} 0.3$ when ${\varepsilon}_0^2
\to 0$. On the other hand,
$\delta ({\varepsilon}^2)_{\mbox{\scriptsize  gap}}$
remains relatively stable as ${\varepsilon}_0^2 \to 0$;
$\delta ({\varepsilon}^2)_{\mbox{\scriptsize  gap}}$ is thus 
identified as the source of the observed
``$1/N_{\mbox{\footnotesize c}}$-nonperturbative'' behavior,
unlike $\delta ({\varepsilon}^2)_{\mbox{\scriptsize  ren.}}$.

Finally, the leading part of QCD effects was included. The 
``gap'' part is represented by the contributions coming from the 
diagrams of Fig.~2, where the internal dashed lines represent now 
the gluon propagators (in Landau gauge). The momentum integrals
were regulated by means of a proper-time cut-off
$1/{\Lambda}_{\mbox{\footnotesize f}}^2$ for the quarks and
$1/{\Lambda }_{\mbox{\footnotesize b}}^2$ for the gluons.
The corresponding contribution to $\Xi ^{(1)}$ to be added in
(\ref{Xi1}) was derived in~\cite{cpv}
\begin{equation}
\Xi^{(1;\mbox{\scriptsize gl})}
\left( {\varepsilon}^2;
{\Lambda }_{\mbox{\footnotesize b}}^2/{\Lambda}
_{\mbox{\footnotesize f}}^2;a_{\mbox{\scriptsize gl}}\right) 
= 2 \int_0^{{\Lambda }_{\mbox{\scriptsize b}}^2/{\Lambda}
_{\mbox{\scriptsize f}}^2}
d{\bar p}^2{\bar p}^2
\ln \left[ 1-a_{\mbox{\scriptsize gl}}
 {\cal {J}}_{\mbox{\scriptsize gl}}\left( {\bar p}^2;
 {\varepsilon}^2\right)
\right] \ ,  
\label{Xigl}
\end{equation}
Above, we denoted by $a_{\mbox{\scriptsize gl}}$ 
the QCD coupling parameter:
$a_{\mbox{\scriptsize gl}}=3\alpha _s(m_t)/\pi \approx 0.105$. 
The (proper-time regulated) 2-point Green function 
${\cal {J}}_{\mbox{\scriptsize gl}}$ appearing in 
(\ref{Xigl}) is 
\begin{eqnarray}
{\cal {J}}_{\mbox{\scriptsize gl}}
\left( {\bar p}^2,{\varepsilon}^2\right)  &=&
-\frac 16\left(2\frac{{\varepsilon}^2}{{\bar p}^2}-1\right) 
 {\cal {E}}\left( \frac{{\varepsilon}^2}{{\bar p}^2}\right) 
 +\frac 16\ln {\varepsilon}^2+\frac 29 
\nonumber \\
&&
 -\frac 16\left( \frac{{\bar p}^2}5+{\varepsilon}^2\right) 
 +\frac 14\left( 
 \frac{{\bar p}^4}{140}+\frac{{\bar p}^2{\varepsilon}^2}{15}
 +\frac{{\varepsilon}^4}6 \right) 
 +{\cal {O}}\left( {\bar p}^6,{\varepsilon}^6 \right) 
\ ,  
\label{Jglpt}
\end{eqnarray}
where we denoted by ${\cal {E}}$ the integral 
\begin{equation}
{\cal {E}}\left( w \right) =
\int_0^1 dz \ln \left[ 1+\frac{z(1-z)}w\right] =
-2+\sqrt{\left( 4w+1\right) }
\ln \left[ 
\frac{\sqrt{\left( 4w+1\right) }+1}{\sqrt{\left( 4w+1\right) }-1}
 \right] \ .  
\label{defE}
\end{equation}
We point out that expression (\ref{Xigl}), unlike (\ref{Xi1}),
turns out to be numerically almost equal to its 2-loop approximation
(obtained by the replacement: 
$\ln [ 1 - a_{\mbox{\scriptsize gl}} 
 {\cal {J}}_{\mbox{\scriptsize gl}} ({\bar p}^2, {\varepsilon}^2 )]
\mapsto - a_{\mbox{\scriptsize gl}} 
{\cal {J}}_{\mbox{\scriptsize gl}}({\bar p}^2, {\varepsilon}^2 ) $),
the difference being only a fraction of a percent.

The leading QCD $m_t$-mass renormalization effect comes from the 
2-loop version of the diagrams of Fig.~3, where the dashed line 
is now the gluonic propagator. The proper-time cut-off gives 
(cf.~\cite{cpv}) 
\begin{equation}
\delta ({\varepsilon}^2)_{\mbox{\scriptsize  ren.}}
^{\mbox{\scriptsize QCD}}=
\frac{2}{3} a_{\mbox{\scriptsize gl}}{\varepsilon}_0^2
\left[ 
\ln \left( {\varepsilon}_0^{-2}\right) 
 +\ln \left( {\Lambda }_{\mbox{\footnotesize b}}^2/
{\Lambda}_{\mbox{\footnotesize f}}^2\right) 
 +0.256 \ldots
 +\frac{5{\Lambda}^2_{\mbox{\footnotesize f}}}
{9{\Lambda}^2_{\mbox{\footnotesize b}}}{\varepsilon}_0^2
 +{\cal {O}}({\varepsilon}_0^4)
\right] \ .
\label{kapglr}
\end{equation}
This expression is to be added to (\ref{kappar}) in order to obtain 
the QCD-modified 
$\delta({\varepsilon}^2)_{\mbox{\scriptsize  ren.}}$.

QCD effects give positive contributions to 
$\delta ({\varepsilon}^2)_{\mbox{\scriptsize gap}}$ and to 
$\delta ({\varepsilon}^2)_{\mbox{\scriptsize ren.}}$. The 
contribution to the ``gap'' 
ntl-shift $\delta ({\varepsilon }^2)_{\mbox{\scriptsize gap}}$ 
is by about one order of 
magnitude smaller than the corresponding contribution of the 
scalars to this quantity. On the other hand, the positive QCD 
contribution to 
$\delta ({\varepsilon}^2)_{\mbox{\scriptsize ren.}}$ is 
larger by a factor of 3--5, and it is
comparable to the positive contribution of the 
neutral Goldstone to this quantity. 
Altogether, $\delta ({\varepsilon}^2)_{\mbox{\scriptsize  ren.}}$ 
is still clearly negative under the inclusion of QCD effects, and 
$\delta ({\varepsilon}^2)_{\mbox{\scriptsize gap}}$ 
remains negative and with a substantially larger magnitude than 
$\delta({\varepsilon}^2)_{\mbox{\scriptsize  ren.}}$ (by a factor
of 2 or more). In Table 2,
we display the results analogous to those of Table 1, but now 
these QCD effects are included. Comparing Table 1 and Table 2, we 
see that the inclusion of QCD changes the results rather 
modestly. The basic result remains the same: as long as we demand 
that the leading-$N_{\mbox{\footnotesize c}} $ gap equation have at 
least a qualitatively predictive power, the cut-off 
${\Lambda}_{\mbox{\footnotesize f}}$ 
($\sim {\Lambda }_{\mbox{\footnotesize b}}$) cannot surpass 
${\cal {O}}(1\mbox{ TeV})$.

In conclusion, we mention that other authors have studied ntl-effects
in the TSM and in related frameworks~\cite{hands}-~\cite{lurie}. 
The authors of~\cite{hands} calculated ntl-contributions to
critical exponents of the fields within NJL-type models at the
fixed point, i.e., at the location of the non-trivial zero of 
$\beta$-function, for various dimensions $d$. The implications
of~\cite{hands} in relation to 4-dimensional NJL-type models at 
low energy and with finite cut-off are not clear and would 
deserve investigation.
When concluding the present work, a
somewhat related work by K.~Akama~\cite{akama} came to our 
attention. Akama investigates the ntl-effects by considering the
compositeness condition, i.e., the condition that the 
renormalization constants of a composite scalar field and of its
self-interaction parameter are zero.
He reaches the conclusion that the ntl-effects for 
$N_{\mbox{\footnotesize c}}=3$ are substantially stronger than 
the leading-$N_{\mbox{\footnotesize c}}$ contributions
and lead to physically unacceptable results: negative Higgs mass,
negative $\Phi^4$-coupling, etc. 
Furthermore, Luri\'e and Tupper~\cite{lurie} had earlier
considered the compositeness condition and took into 
account at least some of the effects beyond the 
leading-$N_{\mbox{\footnotesize c}}$, arriving at qualitatively 
the same conclusion as Akama - that 
$1/N_{\mbox{\footnotesize c}}$-expansion diverges.
We note that these three authors treated the
TSM as a renormalizable Yukawa-type model (without gauge bosons)
plus the compositeness condition, similar to (but not identical with) 
the approach of BHL~\cite{bhl}.
Thus, they implicitly assumed large cut-offs $\Lambda$, in the
sense that $\ln \Lambda$-terms would entirely dominate over the 
$\Lambda$-independent parts. 
Consequently, the results of Akama, Luri\'e and Tupper appear to not
contradict the results of the present paper - i.e., that the TSM
can be interpreted at the ntl-level in a straightforward manner
only if $\Lambda={\cal {O}}(1\mbox{ TeV})$, and that it may be 
difficult or impossible to interpret the model if 
$\Lambda > {\cal {O}}(1\mbox{ TeV})$.

\section{Acknowledgment}

This work was supported in part by the Deutsche Forschungsgemeinschaft
and in part by the European Union Project CHRX-CT92-0026.

\newpage

\vspace{4cm}

\oddsidemargin-1.8cm 
\evensidemargin-1.8cm

\begin{table}[h]
\vspace{0.3cm}
\par
\begin{center}
Table 1 \\\vspace{0.5cm} 
\begin{tabular}{|l|l||c|c|c|c||c|c|c|c|}
\hline
$m_t^{\mbox{\scriptsize  ren.}}/m_t^{(0)}$ & 
${\Lambda}_{\mbox{\footnotesize b}}/
 {\Lambda}_{\mbox{\footnotesize f}}$ & 
$m_t^{(0)}/\Lambda_{\mbox{\footnotesize f}} $ & 
$m_t^{\mbox{\scriptsize  ren.}}/\Lambda_{\mbox{\footnotesize f}}$ & 
${\Lambda}_{\mbox{\footnotesize f}}(\mbox{sc})$ & 
${\Lambda}_{\mbox{\footnotesize b}}(\mbox{sc})$ & 
$m_t^{(0)}/\Lambda_{\mbox{\footnotesize f}}$ & 
$m_t^{\mbox{\scriptsize  ren.}}/\Lambda_{\mbox{\footnotesize f}}$ & 
${\Lambda}_{\mbox{\footnotesize f}}(H)$ & 
${\Lambda}_{\mbox{\footnotesize b}}(H)$ \\ 
&  & (sc) & (sc) & [TeV] & [TeV] & ($H$)& ($H$)& [TeV] & [TeV] \\ 
\hline\hline
$1/\sqrt{2}(=0.707)$ & $1/\sqrt{2}$ & 
 -- & -- & -- & -- & 0.320 & 0.226 & 0.79 & 0.56 \\ 
0.5 & $1/\sqrt{2}$ & 0.757 & 0.379 & 0.48 & 0.34 & 0.250 & 0.125 & 
1.44 & 1.02 \\ 
0.333 &$1/\sqrt{2}$& 0.540 & 0.180 & 1.00 & 0.71 & 0.226 & 0.075 &
2.39 & 1.69 \\ 
0.25 &$1/\sqrt{2}$& 0.500 & 0.125 & 1.44 & 1.02 & 0.219 & 0.055 & 
3.28 & 2.32 \\ 
\hline
0.707 & 0.5 & 0.516 & 0.365 & 0.49 & 0.25 & 0.200 & 0.141 & 
1.27 & 0.64 \\ 
0.5 & 0.5 & 0.329 & 0.164 & 1.09 & 0.55 & 0.161 & 0.081 & 
2.24 & 1.12 \\ 
0.333 & 0.5 & 0.282 & 0.094 & 1.91 & 0.96 & 0.147 & 0.049 & 
3.67 & 1.84 \\ 
0.25 & 0.5 & 0.270 & 0.067 & 2.67 & 1.34 & 0.143 & 0.036 & 
5.04 & 2.52 \\ 
\hline
\end{tabular}
\end{center}
\end{table}

\vspace{2cm}

\oddsidemargin-2.7cm 
\evensidemargin-2.7cm

\begin{table}[h]
\vspace{0.3cm}
\par
\begin{center}
Table 2 \\\vspace{0.5cm} 
\begin{tabular}{|l|l||c|c|c|c|}
\hline
$m_t^{\mbox{\scriptsize ren.}}/m_t^{(0)}$ & 
${\Lambda}_{\mbox{\footnotesize b}}/
{\Lambda}_{\mbox{\footnotesize f}}$ & 
$m_t^{(0)}/\Lambda_{\mbox{\footnotesize f}} $ & 
$m_t^{\mbox{\scriptsize  ren.}}/\Lambda_{\mbox{\footnotesize f}}$ & 
${\Lambda}_{\mbox{\footnotesize f}}(\mbox{sc+gl})$ & 
${\Lambda}_{\mbox{\footnotesize b}}(\mbox{sc+gl})$ \\ 
&  &  & (sc+gl) & [TeV] & [TeV] \\ \hline\hline
$1/\sqrt{2}(=0.707)$&$1/\sqrt{2}$& -- & -- & -- & -- \\ 
0.5   &$1/\sqrt{2}$ & 0.641 & 0.321 & 0.56 & 0.40 \\ 
0.333 &$1/\sqrt{2}$ & 0.461 & 0.154 & 1.17 & 0.83 \\ 
0.25  &$1/\sqrt{2}$ & 0.428 & 0.107 & 1.68 & 1.19 \\ \hline
0.707 & 0.5 & 0.466 & 0.329 & 0.55 & 0.27 \\ 
0.5   & 0.5 & 0.288 & 0.144 & 1.25 & 0.63 \\ 
0.333 & 0.5 & 0.248 & 0.083 & 2.18 & 1.09 \\ 
0.25  & 0.5 & 0.237 & 0.059 & 3.03 & 1.52 \\ \hline
\end{tabular}
\end{center}
\end{table}

\oddsidemargin-0.5cm 
\evensidemargin-0.5cm

\newpage

\section{Table and figure captions}

{\footnotesize
\noindent {\bf Table 1}: The quark (fermion) cut-offs 
${\Lambda}_{\mbox{\scriptsize f}}$ 
and the bosonic cut-offs ${\Lambda}_{\mbox{\scriptsize b}}$ 
which result when we impose 
the requirement that the ratio of the next-to-leading $m_t$ 
(i.e., $m_t^{\mbox{\scriptsize  ren.}}$) to the 
leading-$N_{\mbox{\scriptsize c}}$ $m_t^{(0)}$ be 
$1/\sqrt{2}$, $1/2$, $1/3$ and $1/4$, respectively; we have chosen 
the cut-off ratios ${\Lambda}_{\mbox{\scriptsize b}}/
{\Lambda}_{\mbox{\scriptsize f}} = 1/\sqrt{2}$, $1/2$; 
``sc'' indicates quantities for the case when all four scalar 
degrees were taken into account at the next-to-leading (ntl) level; 
``$H$'' indicates quantities when only the physical Higgs degree of 
freedom was taken into account at the ntl level. We identify 
$m_t^{\mbox{\scriptsize ren.}}= 
m_t^{\mbox{\scriptsize phys.}} = 180 \mbox{ GeV}$.
No entries in the first line for the ``sc'' case mean
that $(m_t^{\mbox{\scriptsize  ren.}}/m_t^{(0)}) < 0.707$
for any choice of $m_t^{(0)}/{\Lambda}_{\mbox{\scriptsize f}}$
(when ${\Lambda}_{\mbox{\scriptsize b}}/
{\Lambda}_{\mbox{\scriptsize f}} = 0.707$).

\vspace{1cm}

\noindent {\bf Table 2}: Same as Table 2, but this time for the 
case when, in addition, the leading part of the QCD (two loop) 
effects was taken into account.

\vspace{1cm}

\noindent {\bf Figs.~1(a)-(c)}: The 1-loop 1-PI diagrams 
contributing to 1-PI Green functions 
$\tilde \Gamma_{H} ^{(2m; 1)}(p_1, \ldots, p_{2m})$,
which in turn yield the leading-$N_{\mbox{\scriptsize c}}$ part 
$V^{(0)}_{\mbox{\scriptsize eff}}$ in the 
formal $1/N_{\mbox{\scriptsize c}}$-expansion of 
$V_{\mbox{\scriptsize eff}}$. Full lines represent massless top 
quarks, and dotted external lines the scalar non-dynamical Higgs 
of the Lagrangian (\ref{TSM1}).

\vspace{1cm}

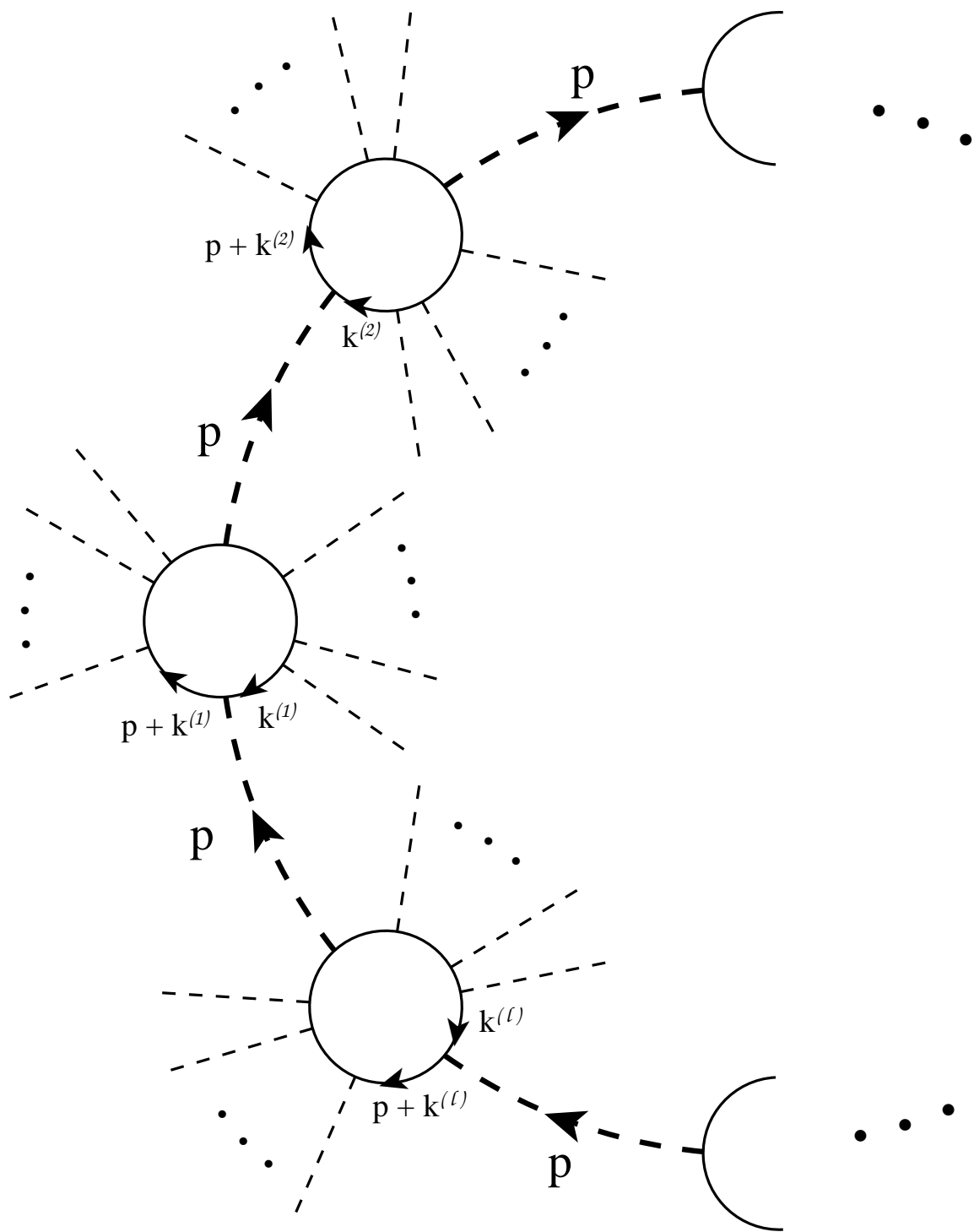
\noindent {\bf Fig.~2}: The $(\ell + 1)$-loop 1-PI diagrams which 
contribute to the 1-PI Green functions which in turn yield the 
ntl-part $V_{\mbox{\scriptsize eff}}^{(1)}$ (beyond 1-loop) in the 
formal $1/N_{\mbox{\scriptsize c}}$-expansion of 
$V_{\mbox{\scriptsize eff}}$. The diagrams contain $\ell$ loops 
of (massless) quarks. These loops are connected into another circle 
by $\ell$ propagators of the (non-dynamical) scalars (all either 
Higgs, or neutral Goldstone, or charged Goldstone). In the case of 
charged Goldstone propagators, the quark loops are made up of the 
top and the bottom quark.

\vspace{1cm}

\noindent {\bf Figs.~3(a)-(c)}: the 1-PI diagrams with two external 
top quark legs which give the leading 
(${\cal {O}}(1/N_{\mbox{\scriptsize c}})$) contribution to the 
renormalization of the mass $m_{t} $. Unlike the diagrams of 
Figs.~1-2, the top quark propagators here contain the non-zero
bare mass $m_{t} $ which was the solution to the 
leading-$N_{\mbox{\scriptsize c}}$ gap 
equation. The dashed lines are all the non-dynamical scalars 
(either the Higgs, or the neutral Goldstones, or the charged 
Goldstones). For the case of charged Goldstone propagators, the 
loops contain one massive top quark and one massless bottom quark.
}

{\small

\begin{thebibliography}{9}

\bibitem{nambu} Y.~Nambu, Proceedings of the Kazimierz Conference
``New Theories in Physics'' (1988), pp.~1-10; V.A.~Miransky,
M.~Tanabashi and K.~Yamawaki, Mod.~Phys.~Lett.~A 4 (1989) 1043; 
Phys.~Lett.~B 221 (1989) 177.

\bibitem{marciano}
W.J.~Marciano, Phys.~Rev.~Lett.~62 (1989) 2793;
Phys.~Rev.~D 41 (1990) 219.

\bibitem{bhl}  W.~A.~Bardeen, C.~T.~Hill and M.~Lindner, 
Phys.~Rev.~D 41 (1990) 1647.

\bibitem{justin} 
M.~Bando, T.~Kugo, N.~Maekawa, N.~Sasakura and Y.~Watabiki, 
Phys.~Lett.~B 246 (1990) 466;
A.~Hasenfratz, P.~Hasenfratz, K.~Jansen, J.~Kuti and Y.~Shen,
Nucl.~Phys.~B 365 (1991) 79; 
S.F.~King and S.H.~Mannan, Phys.~Lett.~B 
241 (1990) 249; Z.~Phys.~C 52 (1991) 59;
C.T.~Hill, M.A.~Luty and E.A.~Paschos, Phys.~Rev.~D 
43 (1991) 3011; Y.~Achiman and A.~Davidson, Phys.~Lett.~B 
261 (1991) 431; F.~Cooper and J.~P\'erez-Mercader, 
Phys.~Rev.~D 43 (1991) 4129; 
J.~Zinn-Justin, Nucl.~Phys.~B 367 (1991) 105; 
P.~Fishbane, R.E.~Norton
and T.N.~Truong, Phys.~Rev.~D 46 (1992) 1768; 
P.~Fishbane and R.E.~Norton, Phys.~Rev.~D 48 (1993) 4924; 
A.~Blumhofer, R.~Dawid and M.~Lindner, Phys.~Lett.~B 360 (1995)
123.

\bibitem{njl} Y.~Nambu and G.~Jona-Lasinio, Phys.~Rev.~122 (1961)
345.

\bibitem{cpv}  G.~Cveti\v c, E.A.~Paschos and N.D.~Vlachos, 
Phys.~Rev.~D 53 (1996) 2820.

\bibitem{kugo}
T.~Kugo, Prog.~Theor.~Phys.~55 (1976) 2032;
K.~Kikkawa, {\it ibid.} 56 (1976) 947.

\bibitem{weinberg} S.~Weinberg, Phys.~Rev.~D 7 (1973) 2887.

\bibitem{hands}
S.~Hands, A.~Koci\'c and J.B.~Kogut, Phys.~Lett.~B 273 (1991) 
111; Ann.~Phys.~224 (1993) 29;
J.A.~Gracey, Phys.~Lett.~B 308 (1993) 65; Z.~Phys.~C 
61 (1994) 115; S.E.~Derkachov, N.A.~Kivel, 
A.S.~Stepanenko and A.N.~Vasil'ev, 
in ``{\it Hadrons-93}\ '', Conference proceedings, (publisher: 
Novy Svit, 1993); Saclay preprint SACLAY-SPHT-93-016 
(hep-th/9302034).

\bibitem{akama}
K.~Akama, Phys.~Rev.~Lett.~76 (1996) 184.

\bibitem{lurie}
D.~Luri\'e and G.B.~Tupper, Phys.~Rev.~D 47 (1993) 3580.

\end{thebibliography}

}
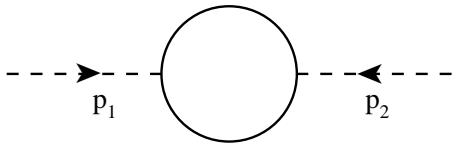
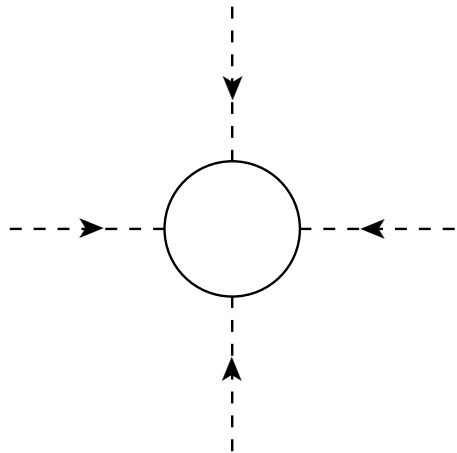
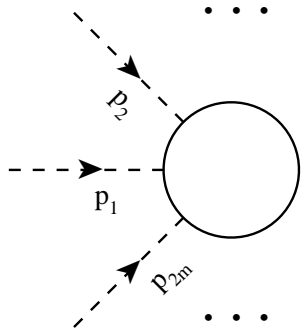
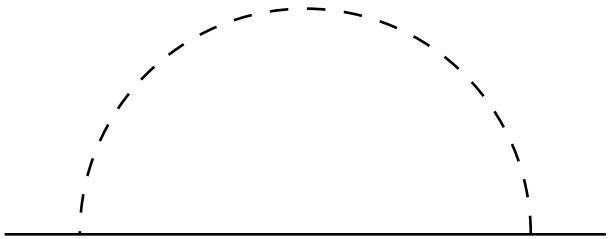
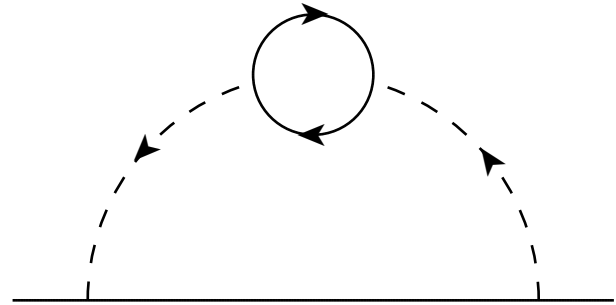
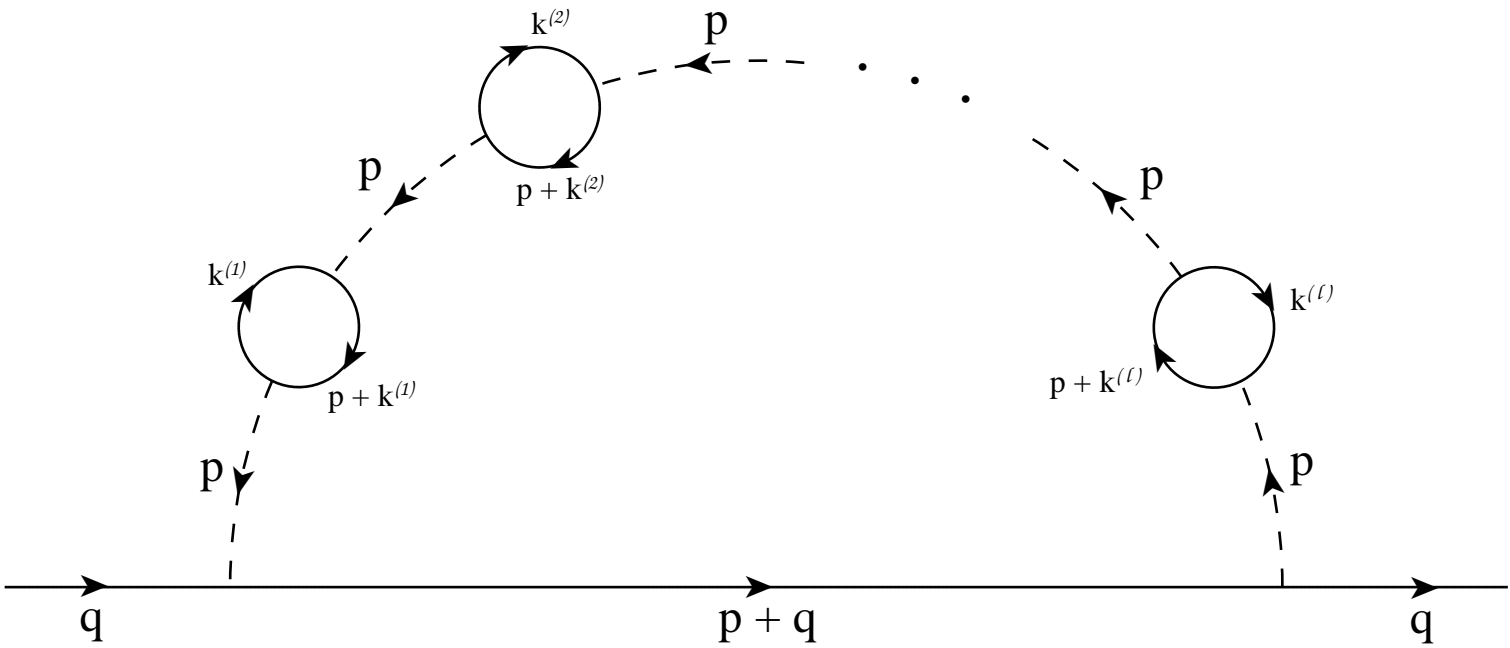
\end{document}